# Strongly Gapped Topological Surface States on Protected Surfaces of Antiferromagnetic MnBi$_4$Te$_7$ and MnBi$_6$Te$_{10}$

Kyle N. Gordon[1], Hongyi Sun[2], Chaowei Hu[3], A. Garrison Linn[1], Haoxiang Li[1], Yuntian Liu[2], Pengfei Liu[2], Scott Mackey[3], Qihang Liu[2,4‡], Ni Ni[3*], Dan Dessau[1,5†]

[1]Department of Physics, University of Colorado, Boulder, CO 80309, USA
[2]Shenzhen Institute for Quantum Science and Technology and Department of Physics, Southern University of Science and Technology, Shenzhen, 518055, China
[3]Department of Physics and Astronomy and California NanoSystems Institute, University of California, Los Angeles, CA 90095, USA
[4]Guangdong Provincial Key Laboratory for Computational Science and Material Design, Southern University of Science and Technology, Shenzhen 518055, China
[5]Center for Experiments on Quantum Materials, University of Colorado, Boulder, CO 80309, USA

† dessau@colorado.edu
* nini@physics.ucla.edu
‡ liuqh@sustech.edu.cn

## ABSTRACT

The search for materials to support the Quantum Anomalous Hall Effect (QAHE) have recently centered on intrinsic magnetic topological insulators (MTIs) including MnBi$_2$Te$_4$ or heterostructures made up of MnBi$_2$Te$_4$ and Bi$_2$Te$_3$. While MnBi$_2$Te$_4$ is itself a MTI, most recent ARPES experiments indicate that the surface states on this material lack the mass gap that is expected from the magnetism-induced time-reversal symmetry breaking (TRSB), with the absence of this mass gap likely due to surface magnetic disorder. Here, utilizing small-spot ARPES scanned across the surfaces of MnBi$_4$Te$_7$ and MnBi$_6$Te$_{10}$, we show the presence of large mass gaps (~ 100 meV scale) on both of these materials when the MnBi$_2$Te$_4$ surfaces are buried below one layer of Bi$_2$Te$_3$ that apparently protects the magnetic order, but not when the MnBi$_2$Te$_4$ surfaces are exposed at the surface or are buried below two Bi$_2$Te$_3$ layers. This makes both MnBi$_4$Te$_7$ and MnBi$_6$Te$_{10}$ excellent candidates for supporting the QAHE, especially if bulk devices can be fabricated with a single continuous Bi$_2$Te$_3$ layer at the surface.

**Introduction**

Magnetic Topological Insulators (MTI) have become the subject of a great deal of recent research effort due to the exciting predicted properties of their topological quantum states, for example the Quantum Anomalous Hall Effect (QAHE) which is a macroscopic quantum phenomenon predicted more than 30 years ago [1] that supports dissipation-less conducting edge channels. Unfortunately, Haldane's original proposal of a honeycomb lattice with zero net magnetic flux has not produced the sought-after QAHE; however, an alternative perspective [2] has led to fruitful experimental work. The proposal stipulates the existence of two conditions: 1) band inversion and 2) ferromagnetic ordering. There already exist many topological insulators (TI), but many of these are time reversal invariant and therefore show zero Hall conductance (the Quantum Spin Hall Effect). The second ingredient, ferromagnetic ordering, breaks time reversal symmetry and couples to one direction of circulation, akin to the external magnetic field required in the Hall Effect. In the absence of time reversal symmetry breaking (TRSB), the surface states would be massless (linear dispersing) ungapped Dirac cones, but with the TRSB induced by the ferromagnetism, these Dirac cones acquire a mass term, i.e. they become gapped [2]. Interestingly, these surface states are further distinguished from massless states by their winding numbers, which produce a quantized half Hall conductance, $e^2/2h$ per level [3]. When arranged in a ferromagnetic configuration this is predicted to lead to the QAH state including its dissipation-less edge currents. Following its theoretical prediction [4, 5, 6], this effect was first observed [7] in the magnetically-doped topological insulator $(Cr_xBi_{1-x})_2Se_3$. An important note is that the effect was only observed at very low temperatures -- of order 100 mK [7]. It is believed that the low-temperature limitation originates from the disorder introduced by the randomly positioned magnetic dopants. For this reason, an intrinsic magnetic topological insulator without the disorder of random dopants has been sought, with $MnBi_2Te_4$ a recently realized prime candidate [8, 9]. This material has spontaneous antiferromagnetic (AFM) magnetization [10, 11, 12] and because manganese is one component of the crystal it does not suffer from the disorder introduced by random doping. Upon the application of a magnetic field to align the spin moments, the quantized states have been observed in bulk crystals [13], though only at very high fields of ~ 5 Tesla and still at relatively low temperatures of 1.6 K. However, more recent work on specially exfoliated nano-devices [14, 15] have reached record high temperatures of 60 K. These high magnetic saturation fields are not ideal for applications such as QAHE, and the overall tunability of the material properties of MnBi$_2$Te$_4$ remains very limited. Another critical issue is that while original Angle Resolved Photoemission Spectroscopy (APRES) measurements of the surface states of MnBi$_2$Te$_4$ indicated that the surface states were indeed gapped [16, 17] in agreement with theoretical prediction [2, 18] and Density Functional Theory (DFT) calculations, later ARPES measurements have all shown massless (gapless) surface states [19, 20]. Presently, this is believed to be due to magnetic disorder at the surface of the material, which contrasts with the A-type AFM ordering in the bulk [19]. Finding a materials system with gapped topological surface states and with low magnetic saturation fields or a ferromagnetic ground state is thus very important.

MnBi$_4$Te$_7$ and MnBi$_6$Te$_{10}$ are bulk heterostructures consisting of stacks of magnetic MnBi$_2$Te$_4$ septuple layers (SL) and non-magnetic Bi$_2$Te$_3$ quintuple layers (QL) [21, 22, 23]. In MnBi$_4$Te$_7$, the heterostructure alternates between MnBi$_2$Te$_4$ and Bi$_2$Te$_3$ layers. In MnBi$_6$Te$_{10}$, the heterostructure alternates between one MnBi$_2$Te$_4$ layer and two Bi$_2$Te$_3$ layers. Supplementary Figure S1 shows transport and magnetization measurements that demonstrate the bulk AFM ordering for MnBi$_6$Te$_{10}$ [23]. Thanks to the Bi$_2$Te$_3$ QLs as buffer layers, the AFM exchange coupling between two adjacent MnBi$_2$Te$_4$ SLs, and thus the saturation fields are significantly reduced, rendering these systems an ideal platform to investigate the QAHE as well as other topological quantum phases [24]. We also present ARPES measurements which directly probe the bulk and surface electronic structure for each possible surface termination in MnBi$_6$Te$_{10}$. We refer the reader to [22] for similar measurements on MnBi$_4$Te$_7$. We also present a comparison with theoretical DFT slab calculations to test our measurement results against theory.

Figure 1 shows a compilation of ARPES data on both MnBi$_4$Te$_7$ (an alternating stack of individual MnBi$_2$Te$_4$ SLs and Bi$_2$Te$_3$ QLs) and MnBi$_6$Te$_{10}$ (an alternating stack of individual MnBi$_2$Te$_4$ SLs with double Bi$_2$Te$_3$ QLs). The spectra show E vs. k dispersions with crossing points or gaps in the energy range of 0.3 to 0.4 eV that indicate that these materials are n-type doped, consistent with most other bismuth selenide and telluride-based materials. We will call this energy range of 0.3 to 0.4 the charge neutrality energy. By using a small-spot (~ 20 micron) ARPES beam (see methods) and scanning the beam spot across the cleaved surfaces, we were able to isolate two different types of ARPES spectra for $MnBi_4Te_7$ and three for $MnBi_6Te_{10}$ that we attribute to the different types of cleavage terminations that are possible for each material. Our assignment of the different terminations for $MnBi_4Te_7$ is shown in the top left panels and was already described in [22]. For the three surfaces of $MnBi_6Te_{10}$ we note that the experimental Fermi Surface (FS) and E vs. k dispersion of panels e and j are very similar to that of pristine $Bi_2Te_3$ [25], so we assign these spectra to the surface with two QL's on the surface, which we term $QL_2$. The Fermi surface of panel c looks most similar to the Fermi surface of the SL surface of $MnBi_4Te_7$ (panel a), so we assign that the SL termination. The final termination, panels d and i, is intermediate between that of panels c and e, so we assign that to the $QL_1$ termination.

We see that the spectra that arise from a surface terminated by one $Bi_2Te_3$ QL (panels g and i) are distinctly different from those that are terminated by either no QL layers (panels f and h) or two QLs (panel j), in that only those that are terminated by a single QL layer display evidence of a gapped state near the charge neutrality energy, whereas the other spectra show non-gapped or weakly gapped (up to 10 meV scale [22]) spectra. Following the recent ARPES literature on the $MnBi_2Te_4$ compounds [19], we argue that the lack of a gap in the SL-terminated spectra (panels f and h) should be due to magnetic disorder near the surface, which we schematically illustrate in the stacks shown above panels a and c. Therefore, the presence of the gap in the QL or $QL_1$-terminated spectra indicates that the magnetic order of the outer-most $MnBi_2Te_4$ SLs is protected by the QL surface termination, whereas the effect of the magnetism at the surface for the double QL surface termination $QL_2$ is too weak to support a significant surface gap.

Figure 2 shows a zoom-in of the critical region near the charge neutrality point for the SL and $QL_1$ terminated surfaces of $MnBi_6Te_{10}$. It is observed that there is a very weak feature dispersing inside the gap of the $QL_1$ surface terminated spectra (panel c), which we analyze in detail by taking multiple MDC (Momentum Distribution Curve) cuts through the spectra within the red box, as shown in panel d. A similar process is carried out for the SL terminated spectra, with each of the MDCs fit to two Lorentzian peaks, with peak centroids shown by the red and blue dots. Panel e shows an overlay of the red (from the $QL_1$ terminated surface) and blue dots (SL terminated surface), indicating essentially identical peak dispersions. We therefore attribute the weak dispersive features within the gap of the $QL_1$ surface as a cross-contamination of spectra from SL-terminated surfaces, that may even be well below 1 micron-scale or less, i.e. not resolvable with present high-resolution ARPES instrumentation. Similarly, one could consider the possibility of weak contamination from Bi2Te3 intergrowth, which would produce a spectrum similar to, but slightly different[1] than, Fig. 1(j). The very recent ARPES data on these compounds either did not consider the possibility of multiple surface terminations [26] or of cross-contamination between the surfaces [27], and so did not discuss these strongly gapped surface states.

Figure 3 shows another view of the $QL_1$ terminated spectra from both $MnBi_4Te_7$ and $MnBi_6Te_{10}$, including an EDC (Energy Distribution Curve) taken along the k=0 line (Γ point) for both. The EDCs (red curves) are fit using a superposition of Voigt-profile lineshapes (green) to produce the blue curves, which match the experimental (red) curves very closely. The Voigt peaks above and below the gapped region are separated by 100 meV and 140 meV for $MnBi_4Te_7$ and $MnBi_6Te_{10}$, respectively, though the tails of each of these peaks will fill in the gapped regions and give rise to much smaller energy windows of very low spectral weight. These regions are of the most interest for the QAHE as they will have the least possible amounts of contamination from other states, such as those from the bulk electronic structure. Among the presently available materials, our results show that $MnBi_6Te_{10}$ is especially interesting because of its large gap, especially if it could be fabricated with layer-by-layer MBE growth so that the entire sample surface has the preferred $QL_1$ termination. Also helpful will be to bring the chemical potential to the charge neutrality point, for example by replacing small amounts of Bi with Sb, as has been successfully done for other Bi chalcogenide topological insulators [28].

To understand the topological nature of AFM $MnBi_6Te_{10}$ and to interpret the ARPES spectra we measured, we perform both bulk and surface slab DFT calculations. Similar to $MnBi_2Te_4$, $MnBi_6Te_{10}$ crystallizes in the space group *R-3m*. Taking into account the A-type

AFM, the time-reversal symmetry $T$ is broken, however, a combined symmetry $S = T\tau_{1/2}$ is preserved where $\tau_{1/2}$ is the half translation along the $c$ axis of the AFM primitive cell, leading to a $Z_2$ topological invariant defined at the Brillouin-zone plane with $\mathbf{k} \cdot \tau_{1/2} = 0$. In our calculation, we define a primary unit cell with its c vector along $\tau_{1/2}$, thus the $Z_2$ number here is defined on the $k_z = 0$ plane. Supplementary Fig. S2 shows the calculated band structure of bulk AFM MnBi$_6$Te$_{10}$ with the presence of spin-orbit coupling. The conduction band minimum is located at the $\Gamma$ point, while the valence band maximum in the vicinity of $\Gamma$ shows a slightly curved feature. The calculated bulk band gap is about 150 meV. The projection of band eigenstates onto the $p$-orbitals of Bi and Te indicates the band inversion at the $\Gamma$ point, which is consistent with the nontrivial topological nature. By employing the Wilson loop method, we indeed find a nontrivial $Z_2=1$, confirming that MnBi$_6$Te$_{10}$ is indeed a $Z_2$ AFM topological insulator.

Compared with TIs with $T$-symmetry, the protection of gapless surface states in AFM TIs is more fragile, depending on the magnetic moments at the cleaved surface. Assuming a perfect A-type AFM configuration at the surface, all three types of surface terminations SL, QL$_1$ and QL$_2$ should be gapped because of the broken $S$ symmetry, which is inconsistent with our measurements. Therefore, we build slab models to take the magnetic disorder at the surface into account and compare them with our ARPES data. The slab calculations, which utilize periodic boundary conditions, can approximate a surface termination by inserting a vacuum region. Figure 4 shows the DFT slab calculations for SL, QL$_1$ and QL$_2$ terminations of MnBi$_6$Te$_{10}$ near the charge neutrality region. In order to obtain the surface states of a given termination, we choose symmetric slabs with the same top and bottom terminations, and thus constructed the slabs with a thickness of 10, 12 and 8 vdW-layers for the SL, QL$_1$ and QL$_2$ surface calculations, respectively. In order to simulate the possible paramagnetic SL surface, we fix the magnetic moment of the Mn atoms to zero while keeping the buried Mn atoms with A-type AFM. As shown in Fig. 4(a), the dispersion with an X-shape surface state is qualitatively consistent with our experimental findings, but with a gap of 35 meV. This is from the residual magnetic moment at the surface layer after self-consistent calculation and the finite size effect, i.e., the hybridization between the top and bottom surfaces. The calculated result shown in Fig. 4(b) shows a finite gap near charge neutrality, because of the A-type AFM spin orientation protected by the surface QL layer, consistent with our experimental findings. For the same reason, the $QL_2$ termination is also expected to have a gap, while our DFT calculation suggests a metallic feature, as shown in Fig. 4(c). This is because the surface potential drags the surface conduction band down, merging into the valence band, rather than producing a topologically protected Dirac cone. Therefore, by tuning the surface potential, one should be able to restore the expected gapped feature.

To summarize, we have studied the electronic structure of both bulk and surface states in $MnBi_6Te_{10}$ with ARPES experiments and with theoretical DFT calculations. We separated and identified ARPES spectra due to each possible surface termination, and our analysis confirms the presence of a gapped topological surface state. We have further confirmed that $MnBi_6Te_{10}$ is a nontrivial topological insulator with AFM ordering, which is to be expected given its resemblance to $MnBi_4Te_7$. However, we have shown that $MnBi_6Te_{10}$ exhibits a larger topological surface gap than $MnBi_4Te_7$, which may make the material more convenient for observation of QAHE. Furthermore, we have shown that exactly one $Bi_2Te_3$ layer protects the magnetic ordering and opens a surface gap, but an additional $Bi_2Te_3$ layer closes the gap.

**Methods**

**Sample growth and characterization.** We have grown single crystals of $MnBi_{2n}Te_{3n+1}$ ($n$ = 2 and 3) using self-flux [**Error! Bookmark not defined.**]. Mn, Bi, and Te elements are mixed so the molar ratio of MnTe: $Bi_2Te_3$ is 15:85. In each growth, a few sizable plate-like $MnBi_4Te_7$ and $MnBi_6Te_{10}$ single crystals were obtained. Extra care was paid in screening out the right piece. X-ray diffraction at low angles for both the top and bottom (0 0 $l$) surfaces as well as powder X-ray diffraction were performed on a PANalytical Empyrean diffractometer (Cu Ka radiation).

**ARPES measurements.** ARPES measurements on single crystals of $MnBi_6Te_{10}$ were carried out at the Stanford Synchrotron Research Laboratory (SSRL) beamline 5-2 with photon energies between 26 and 36 eV with linear horizontal polarization. Single crystal samples were top-posted on the (001) surface and cleaved in-situ in an ultra-high vacuum better than $4 \times 10^{-11}$ Torr and a temperature of 15 K. ARPES spectra were taken at 12 K, slightly higher than 11 K, the Neel temperature. As the cleaved terrain is expected to consist of patches of exposed [$Bi_2Te_3$] QL layers and [$MnBi_2Te_4$] SL layers, to eliminate the effect of possible QL and SL mixing on the ARPES data, we scanned a 1 mm square surface of the sample in 50 um steps with a 50 um beam spot and collected spectra from over 200 different spots on the sample. We examined each spectrum, finding many regions with clear, sharp features. We also narrowed the beam spot down to 20 x 20 um and scanned more finely, in 15 um steps, in smaller regions of interest. We found that there were regions on the order of 50 x 50 um that were spectroscopically stable, meaning the ARPES spectra were not changing from spot to spot. We took our data with a 20 x 20 um beam spot and studied the centroid of the spectroscopically stable regions, which we believe will minimize any contamination due to another surface.

**First-principles calculations.** We apply density functional theory (DFT) by using the projector-augmented wave (PAW) pseudopotentials [29] with the exchange-correlation of Perdew-Burke-Ernzerhof (PBE) form [30] and GGA+U [31] approach within the Dudarev scheme as implemented in the Vienna ab-initio Simulation Package (VASP) [32, 33, 34, 35]. The energy cutoff is chosen 1.5 times as large as the values recommended in relevant pseudopotentials. The U value is set to be 5 eV[6]. The $k$-points-resolved value of BZ sampling is $0.02 \times 2\pi/\text{Å}$. The total energy minimization is performed with a tolerance of $10^{-6}$ eV. The crystal structure and atomic position are fully relaxed until the atomic force on each atom is less than $10^{-2}$ eV Å. SOC is included self-consistently throughout the calculations. We constructed the slabs with thickness of 10, 12 and 8 vdW-layers for SL, $QL_1$ and $QL_2$ surface calculations, respectively. We constructed Wannier representations [36, 37] by projecting the Bloch states from the DFT calculations of bulk materials onto the Mn-3$d$, Bi-6$p$ and Te-5$p$ orbitals. The Wilson loop method is used for $Z_2$ calculation as implemented in the WannierTools Package.

**Acknowledgments**
We thank Makoto Hashimoto and Donghui Liu at SSRL for experimental help. Work at CU Boulder was supported by NSF-DMR 1534734, work at UCLA was supported by the U.S. Department of Energy (DOE), Office of Science, Office of Basic Energy Sciences under Award Number DE-SC0011978. Work at SUSTech was supported by the NSFC under Grant No. 11874195, and Center for Computational Science and Engineering of Southern University of Science and Technology. HC acknowledges the support from US DOE BES Early Career Award KC0402010 under Contract DE-AC05-00OR22725. Use of the Stanford Synchrotron Radiation Lightsource, SLAC National Accelerator Laboratory, is supported by the U.S. Department of Energy, Office of Science, Office of Basic Energy Sciences under Contract No. DE-AC02-76SF00515.

**Author contributions**
D.D., N. N. and Q. L. supervised the research. K.G., A.G.L., H.L and D. D. carried out the ARPES measurements and data analysis. C. H., J. L., E. E., H. B. and N. N. grew the bulk single crystal and carried out X-ray and transport measurements. Q. L., P. L., H. S. and Y. L. performed the first-principles calculations. H. C., L. D. and C. H. carried out structure determination. D. D, N.N. and K. G. prepared the manuscript with contributions from all authors.

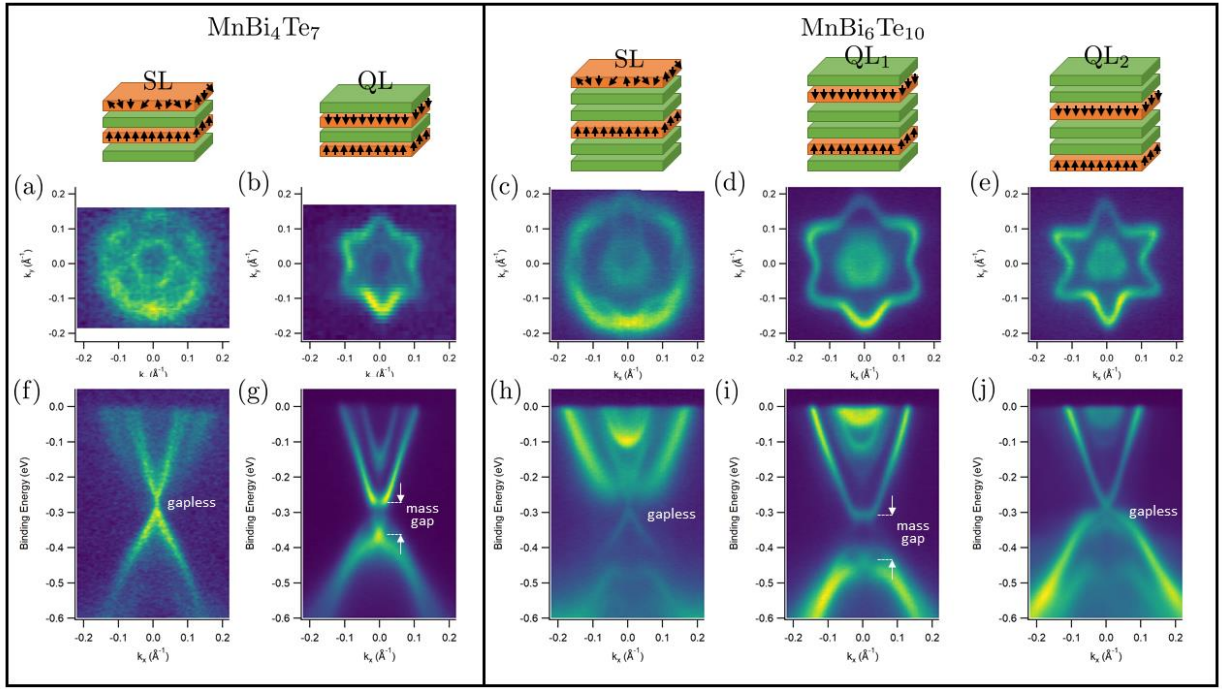

**Figure 1 | Gapped and gapless surface states on MnBi$_4$Te$_7$ and MnBi$_6$Te$_{10}$. a-e** Isoenergy surfaces at the Fermi level for the (**a**) MnBi$_4$Te$_7$ SL, (**b**) MnBi$_4$Te$_7$ QL termination, (**c**) MnBi$_6$Te$_{10}$ SL, (**d**) MnBi$_6$Te$_{10}$ QL$_1$, and (**e**) MnBi$_6$Te$_{10}$ QL$_2$ termination. All Fermi surfaces show the expected six-fold symmetry with clear differences between linearly independent surface terminations. **f-j** $K \to \Gamma \to K$ ARPES measurements for the various terminations, with the charge neutrality point at ~ 0.3 eV (both materials are unintentionally n-doped). Panels (g) and (i) show a gap in the surface states near the charge neutrality point, while the other panels show gapless or near-gapless surface states. The special gapped surfaces correspond to those with one Bi$_2$Te$_3$ QL on the surface.

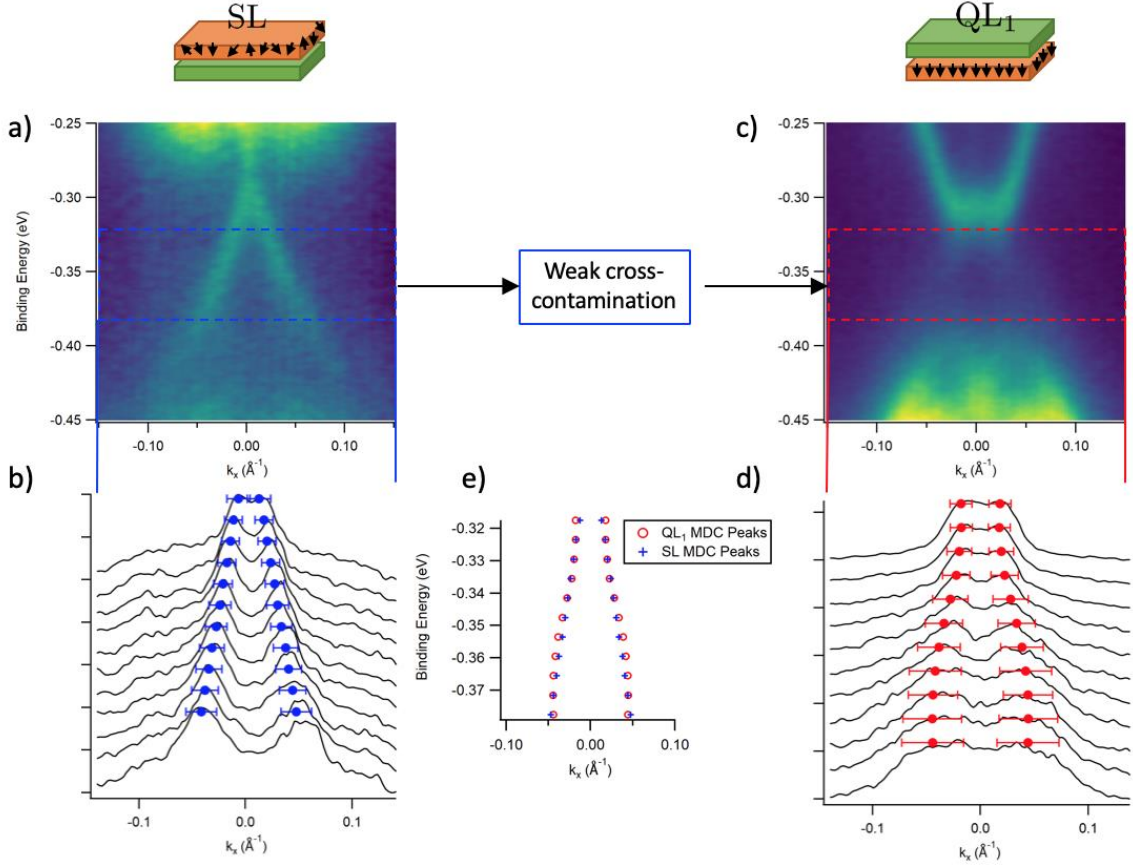

**Figure 2 | Weak cross-contamination in the gapped region of MnBi$_6$Te$_{10}$. a** Zoom-in on a portion of Fig 1h (non-gapped spectra). **b.** The blue dashed rectangle is further examined with MDCs binned in 6 meV windows and stacked vertically. The blue dots show the location of the MDC peaks with the error bars proportional to the full width half max of the MDC peaks. **c** Zoom-in on a portion of Fig 1i (gapped spectra) that shows a weak dispersive band inside the gap. **d.** The red dashed rectangle is further examined with MDCs binned in 6 meV windows and stacked vertically. The red dots show the location of the peaks with error bars proportional to the full width half max of the fitted Lorentzians. **E.** A plot of binding energy versus MDC peak locations, showing the *E-k* dispersion for the weak (red) states in panel d is essentially identical to that from panel b.

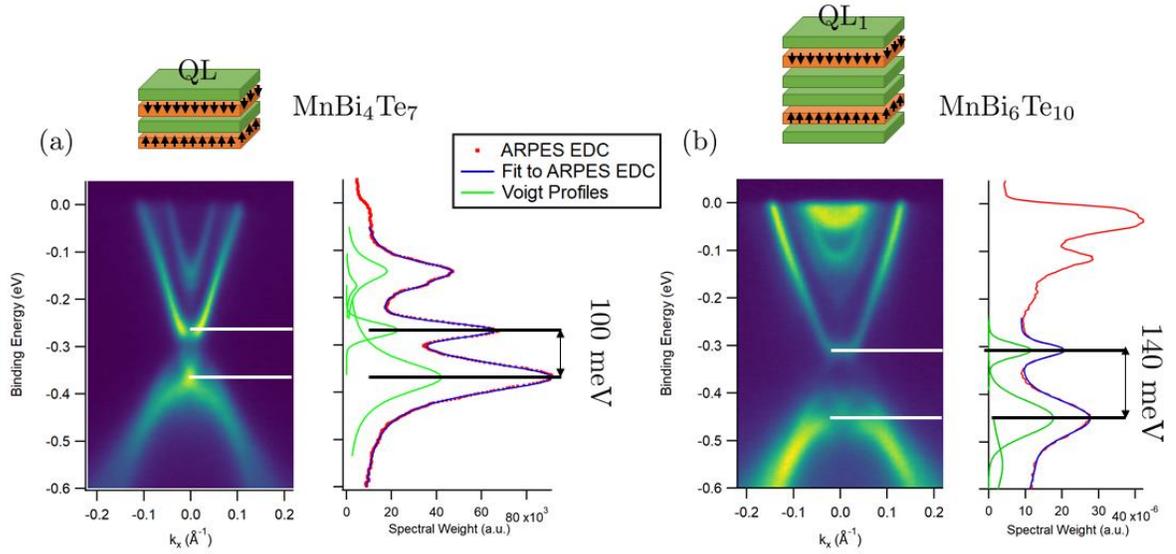

**Figure 3 | A gapped topological surface state in MnBi$_4$Te$_7$ and MnBi$_6$Te$_{10}$. a** QL spectra from MnBi$_4$Te$_7$ (adapted from [**Error! Bookmark not defined.**]) and an EDC at the Γ point. **b** QL$_1$ spectra from Fig. 1 but with an EDC through the Γ point. The red curve shows the experimentally measured EDC, the blue curve is the result of nonlinear curve fitting, and the green curves are the individual Voigt profiles that sum together to create the fitted curve. The difference in peak locations for the relevant Voigt profiles are used to calculate the surface state gap, which are partially filled in by the "tails" of the various states, i.e. the zero-weight gapped region will be considerably smaller.

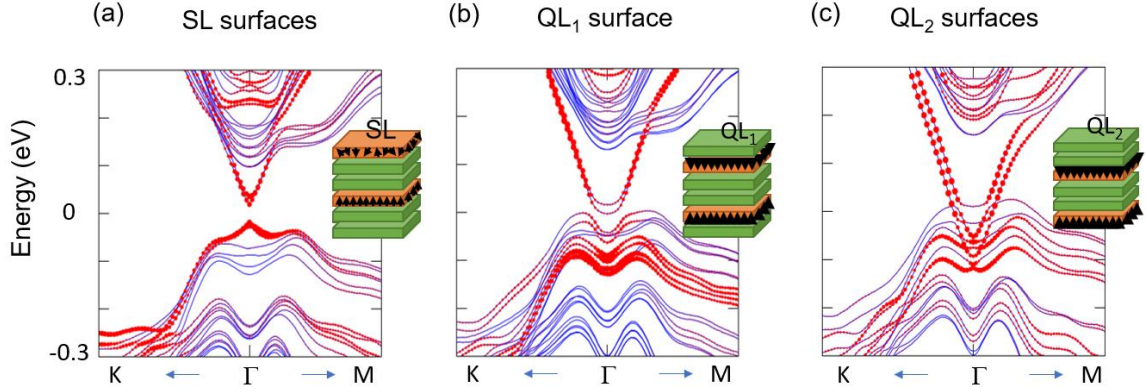

**Figure 4 | Density Functional Theory slab calculations along the z-axis of MnBi$_6$Te$_{10}$ near the charge neutrality region.** The slab calculations, which utilize periodic boundary conditions, can approximate a surface termination by inserting a vacuum region between the two surfaces of the slab. The calculated result in (b) shows a finite gap near charge neutrality, while that in (c) shows a zero or very small gap, qualitatively consistent with our experimental findings. Similar to findings in MnBi$_2$Te$_4$, the finite gap in (a) is inconsistent with experiment and is discussed in the text.

---

1  The relative energy separation between bulk conduction bands and surface bands appears larger than reported results in pristine Bi$_2$Te$_3$ crystals.